\documentclass[aip,reprint]{revtex4-1}
\usepackage{graphicx}
\usepackage{svg}
\usepackage{amsmath}
\usepackage{xcolor}
\draft % marks overfull lines with a black rule on the right

\newcommand{\ket}[1]{$|\,#1\rangle$}

\begin{document}

\title{Entanglement-based clock syntonization for quantum key distribution networks. Demonstration over a 50\,km-long link} %Title of paper

\author{Yoann Pelet}
\email[]{yoann.pelet@unice.fr}
\affiliation{Université Côte d’Azur, CNRS, Institut de Physique de Nice, Nice, France}
\author{Grégory Sauder}
\affiliation{Université Côte d’Azur, CNRS, Institut de Physique de Nice, Nice, France}
\author{Sébastien Tanzilli}
\affiliation{Université Côte d’Azur, CNRS, Institut de Physique de Nice, Nice, France}
\author{Olivier Alibart}
\affiliation{Université Côte d’Azur, CNRS, Institut de Physique de Nice, Nice, France}
\author{Anthony Martin}
\affiliation{Université Côte d’Azur, CNRS, Institut de Physique de Nice, Nice, France}

\date{\today}

\begin{abstract}
We present the implementation of a time synchronization protocol as part of an experimentally deployed entanglement-based quantum key distribution (QKD) link. The system is deployed over $48$\,km of optical fibers across the Métropole Côte d'Azur and enables secret cryptographic key exchange between two remote users, with an average rate of $7$\,kbps. We exploit the time correlation of paired photons generated by a high-quality source of energy-time entanglement implemented in the QKD link to synchronize two rubidium clocks located at the end stations. The level of stability achieved guarantees a time offset between the clocks under $12$\,ps at all time. We also show that this protocol requires less hardware than a typical synchronization protocol for QKD that would distribute a reference clock signal between the users.
\end{abstract}


\maketitle %\maketitle must follow title, authors, abstract and \pacs

\section{Introduction}

Quantum key distribution (QKD) is catching much of the spotlight of quantum technologies, and quickly found its way toward practical applications and commercial development\cite{labonte2024integrated, xu2020secure, pirandola2020advances, rusca2024quantum}.
This has been made possible by the recent progress of QKD technologies in terms of keyrate~\cite{grunenfelder2023fast} and/or link distance between users~\cite{liu2023experimental}, and by the high diversity of protocols available, that can each provide practical solutions to real-field QKD scenarios.
Among those, the most advanced implementation (commercial developments) mostly concerns prepare and measure protocols, while more advanced approaches are still struggling to leave laboratory environments due to the required additional developments.

For example, in entanglement-based QKD systems, time synchronization is one of the most crucial factors, since it allows users (Alice and Bob) to assign an absolute time-stamp to every detection, and to reconstruct time-correlated events.
Precise synchronization can significantly increase signal-to-noise ratio and, taking into account the current performance of single photon detector timing jitter, precision down to tens of picoseconds are now required~\cite{QUANTUM,Wu:17,Chen:13}.
For lab demonstration, a single local time-reference is commonly used, but as soon as real field applications are considered, remote time references at each end station have to be synchronized or syntonized.
Generally synchronization is defined by sharing frequency and phase (i.e. two clocks tick at the same rate and give exactly the same date) whereas syntonization only considers sharing frequency (i.e. two clocks ticks at the same rate and may have a fixed offset).
For many QKD elementary links, syntonisation might be enough, while synchronization is required for advanced multiple-user quantum networking. Whatever the considered application, syntonization can be implemented by a global navigation satellite system (GNSS) at the expense of low accuracy at short timescales~\cite{GNSS_07} or by transmitting intense light pulses, either via a dedicated fiber~\cite{ribezzo2023deploying,WR:22} (White Rabbit technology) or using the quantum channel at a different wavelength~\cite{dynes2019cambridge}, for high accuracy at short timescales. For real--field deployment, the previous solutions might not be compatible with commercial scenarios since wavelength multiplexing adds optical noise. Moreover, exploiting a dedicated fiber for synchronization purposes represents too many additional costs for being seriously considered. The development of standalone, self synchronized, QKD systems is therefore of high interest.
For example, an elegant solution based only on the analysis of detected quantum signal without requiring any additional hardware has been discussed and demonstrated in references~\cite{Qbit4sync_20,Miller_2024,Wang:21}.
It is, however, limited to identify the time of arrival of attenuated laser pulses assuming the repetition rate was known for prepare-and-measure QKD protocols.
Alternatively, for entanglement-based protocols, it is possible to use the strong temporal correlations of the photon pairs transmitted through the network to achieve a very high level of syntonization. 
This method exploits unused information carried through the quantum channel without adding any noise to the qubits, therefore avoiding the drawbacks of other time synchronization methods.
Lastly, this protocol, despite only allowing the syntonization of two clocks, allows to identify exactly correlated events between two users, and will therefore be called "synchronization" in this paper, since the syntonization of the clocks is a byproduct of the photon pair's synchronization.

Thus, we present the implementation of a photon pair based synchronization protocol~\cite{ho2009clock}, exploited over a real-field QKD tesbed deployed over $48$\,km of optical fiber across the Métropole Côte d'Azur~\cite{QUANTUM}.
We show that the achieved level of synchronization allows to generate an optimal keyrate of $7$\,kbps.
Moreover, we push the protocol even further in demonstrating that the two end stations' clocks never drift by more than $12$\,ps.
Lastly, our implementation also passively corrects the daily variations of optical path due to temperature changes in the deployed dark fibers that have to be estimated and rectified when using standard methods.

\begin{figure*}[t]
	\begin{center}
   \includegraphics[width=1.95\columnwidth]{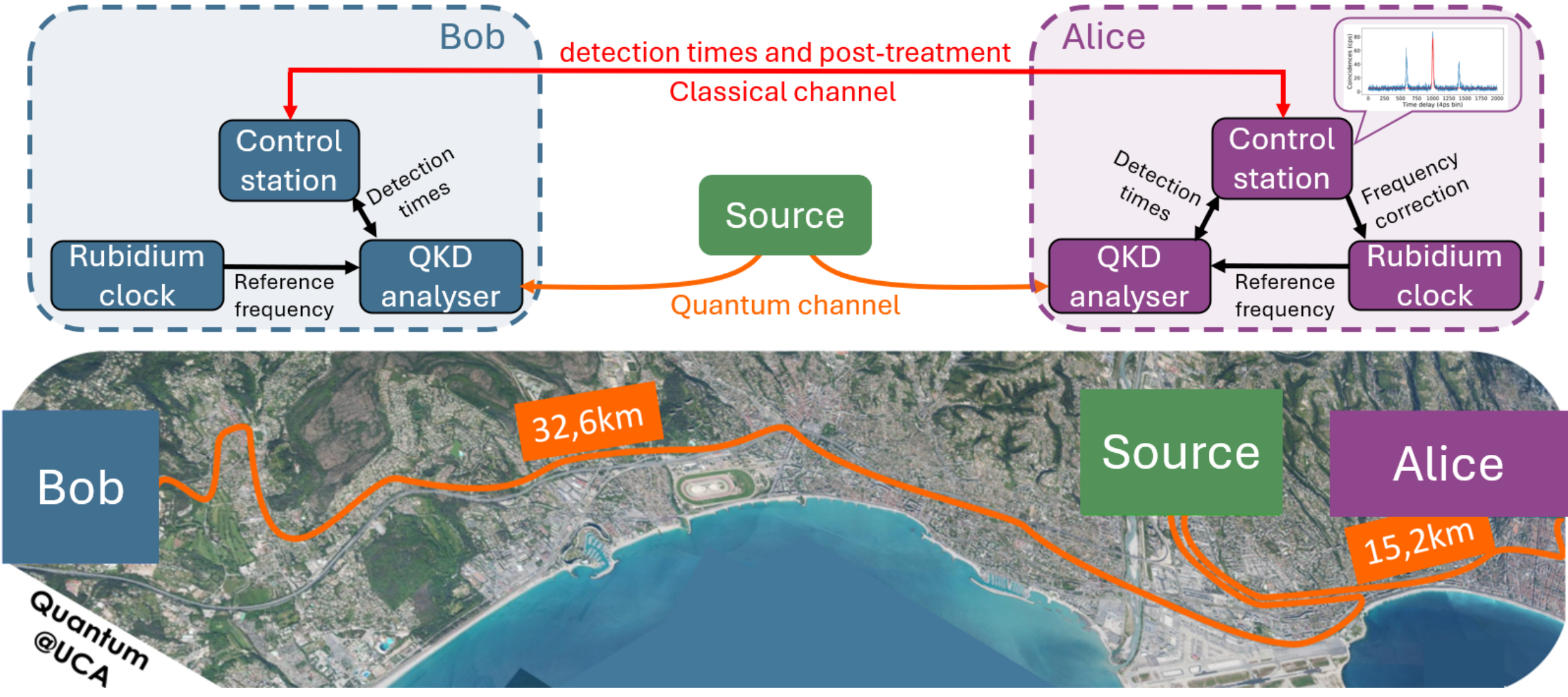}
	\caption{Bottom: general layout of the deployed fibers linking the source to each end station. oranges lines each represent two smf28 optical fibers, one for the quantum signal, and one for the classical communication. Top: scheme of the architecture of the synchronization protocol, where orange and red lines are optical fibers, while black ones are standard USB connections. First, the source generates photon pairs, distributed to each user. Then, Bob sends his detection times to Alice, who create a coincidence histogram by comparing them with hers. This histogram allows for a measurement of the frequency drift between Alice's and Bob's rubidium clock, which is corrected by adapting the frequency of Alice's clock.}
	\label{fig_QKD}
    \end{center}
\end{figure*}

\section{Entanglement based QKD system}
\subsection{Quantum key distribution testbed}
\label{sec_testbed}

The synchronization protocol is implemented over a real-field fully operational QKD link, deployed over $48$\,km of optical telecom fiber\cite{QUANTUM}.
As shown in Fig.\ref{fig_QKD}, the link is composed of three different nodes: two users (Alice and Bob) located at each end station, and the entangled photon pair source, located at the central station. 
The source generates energy-time entangled photon pairs that are deterministically separated with wavelength demultiplexing and distributed to the two end users.
Both stations are equipped with a time bin QKD analyzer, with passive basis choice toward either a Michelson interferometer to project photons on states \ket{+} and \ket{-}, or toward a path encoding setup to measure in the basis \{\ket{0}, \ket{1}\} and ultimately generate the bits of the secure key. 
All details on how this QKD system works can be found in ref\cite{QUANTUM}

The link has been operational for more than a year and can operate for 48 hours without interruption.
Each 48\,h, the single photon detectors need an hour long evaporation cycle, after which the protocol resumes automatically.
The QKD process follows the protocol BBM92~\cite{bennett_quantum_1992} and generates continuously usable cryptographic keys with a fully automated post-treatment program, implemented with LabVIEW.
This control software handles sifting, error correction (implemented as a cascade protocol~\cite{martinez2014demystifying}), privacy amplification, and all required stabilization, including the clock synchronization between the two distant users.

\subsection{Requirement for synchronization}

\begin{figure}[htb]
	\begin{center}
   \includegraphics[width=0.95\columnwidth]{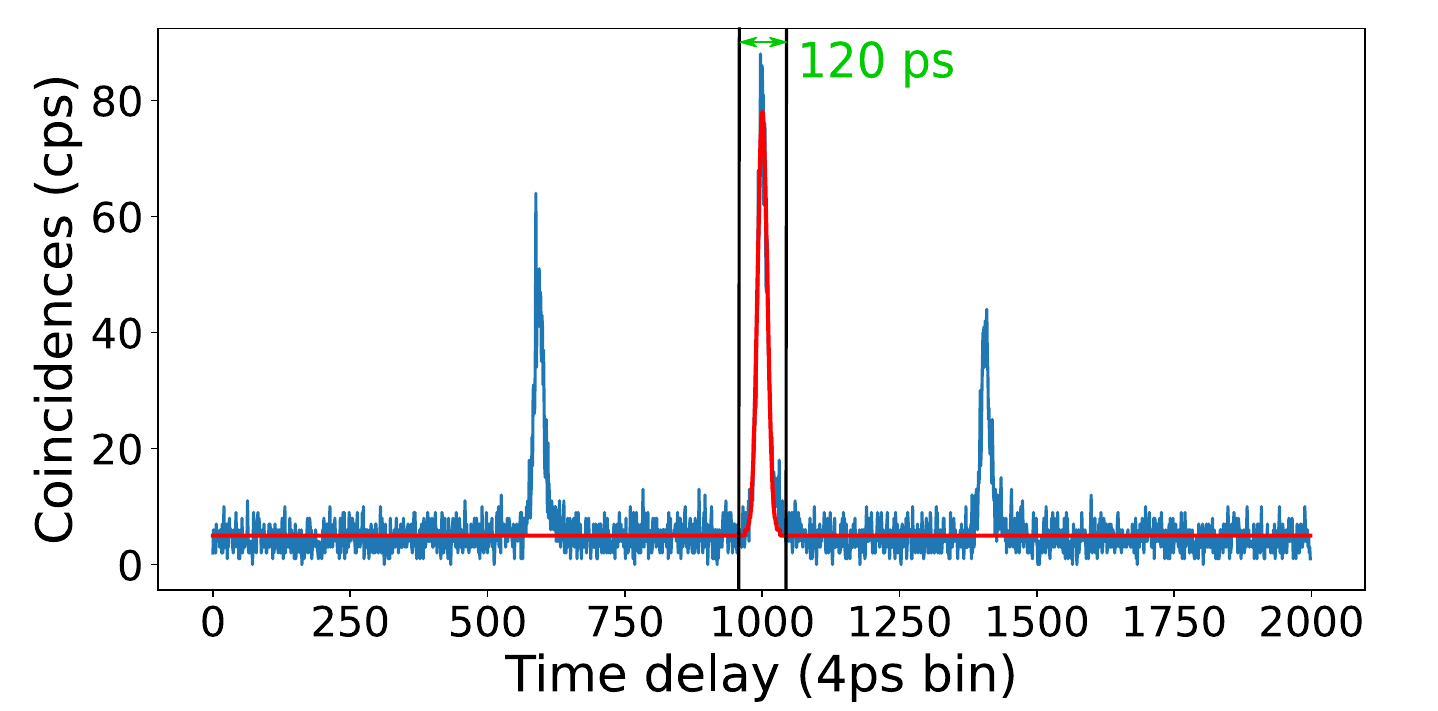}
	\caption{Coincidence histogram plotted twice per second to guarantee the synchronization. The fitted central peak (red) is set at the center of the 120\,ps wide coincidence window (black lines) at each computation of the histogram.}
	\label{fig_3peaks}
    \end{center}
\end{figure}

A critical requirement for entanglement-based QKD is to be able to identify, with a high probability, correlated detections coming from two photons of a same pair, in the presence of noise and losses.
In our protocol, the source generates entangled photon pairs, which are deterministically separated and sent to each end user. 
After the detections of the photons at the end stations, the time-tags registered by Alice are compared to those associated with Bob's detections in a given time window.
This allows to compute the delay histogram between all detections, in which time correlated detections (the photon pairs) form 3 peaks, as shown in Fig.~\ref{fig_3peaks}

In our implementation, the information carried by the photons is projected along four different paths (corresponding to \ket{0}, \ket{1}, \ket{+} and \ket{-}).
Each combination of those paths taken by the paired photons contribute to one of the three peaks in the histogram.
The secured information, however, is only shared by pairs detected in the central peak, which contains the projections on \ket{00}, \ket{11}, \ket{++} and \ket{--}. Therefore, the first challenge for the synchronization protocol is to be able to identify and isolate the central peak.

To estimate the required precision, we determine the width of the peaks, which depends on the dispersion along the fibers, the coherence time of the photons, and the timing jitter of the detection setup. The dispersion along the $50$\,km of fibers is non-locally compensated with a negative dispersion fiber before Alice's analyzer and can therefore be neglected\cite{steinberg1992dispersion}.
The coherence time of the photons is linked to the spectral width of the exploited telecom channels: we use 100\,GHz wide DWDM, resulting in a coherence time of 10\,ps for our photons.
Finally, for the detection setup, Alice and Bob are equipped with superconducting nanowire single photon detectors (SNSPD IDQuantique ID281) showing a timing jitter of 60\,ps.
Timestamps are recorded with two time-to-digital converters (TDC - Swabian Instrument Timetagger Ultra) with a $10$\,ps jitter. 
The combination of the detection setup jitters and coherence time of the photons gives a coincidence peak with a FWHM of $\sim$80\,ps.
To avoid unnecessary noise in the coincidence measurement, we set a coincidence window as narrow as possible, while integrating a maximum of data coming from the peak.
In our implementation, we determine experimentally that the optimal width of this window is $120$\,ps.\\

The second element required to ensure a correct measurement of correlated events is to compensate the natural drift of the clocks.
Indeed, in real implementation, while the position of the coincidence peak ($\delta t_{photons}$) is mostly fixed in time, the coincidence window's position ($\delta t_{clock}$) is affected by the drift of the user's clocks.
As we can only rely on the clocks, every coincidence detected with a delay of $\delta t_{clock} \pm 60$\,ps is considered as part of the central peak and therefore considered as an actual detected pair.
Therefore, if the clock's drift by more than $60$\,ps, at least half of the actual coincidence peak will be ignored since $\delta t_{photons} \notin \{\delta t_{clock} \pm 60\}$.
Comparing the width of the peak and that of the coincidence window, we find that we need $|\delta t_{photons} - \delta t_{clock}|< 30$\,ps in order to keep most of the coincidence peak inside the coincidence window.
We emphasize that this level of synchronization can be achieved and guaranteed by exploiting the photon pairs generated for the QKD protocol, as shown in FIG.\ref{fig_QKD}, where no external reference is directly distributed between the different users.
Before correcting the drift however, the first time offset $\delta t_{photons}$ must be measured.

\section{Global time offset estimation}
\label{sec_offset}

The first measurement required when starting the QKD protocol is to find the time delay between a detection on Alice's side and one on Bob's coming from the same photon pair.
This delay corresponds to the sum of the optical and electrical path difference between Alice and Bob and the time offset between the users' clocks.
The path difference induces a mostly constant offset around $84$\,$\mu$s, owing to the $20$\,km propagation length difference between the two users. 
The clock signal, on the other hand, corresponds to the elapsed time since the initialization of the TDC.
Therefore, if the TDCs are not initialized at the exact same time, a random offset between the two clocks will appear.
Starting two remote TDCs at the same time with a precision on the order of the picosecond is impractical.
We therefore choose to measure this unknown offset instead of trying to minimize it.
Also, since this operation has to be done only once per protocol, we can neglect the frequency drift of the clocks in this section.

\begin{figure}[htb]
	\begin{center}
   \includegraphics[width=0.95\columnwidth]{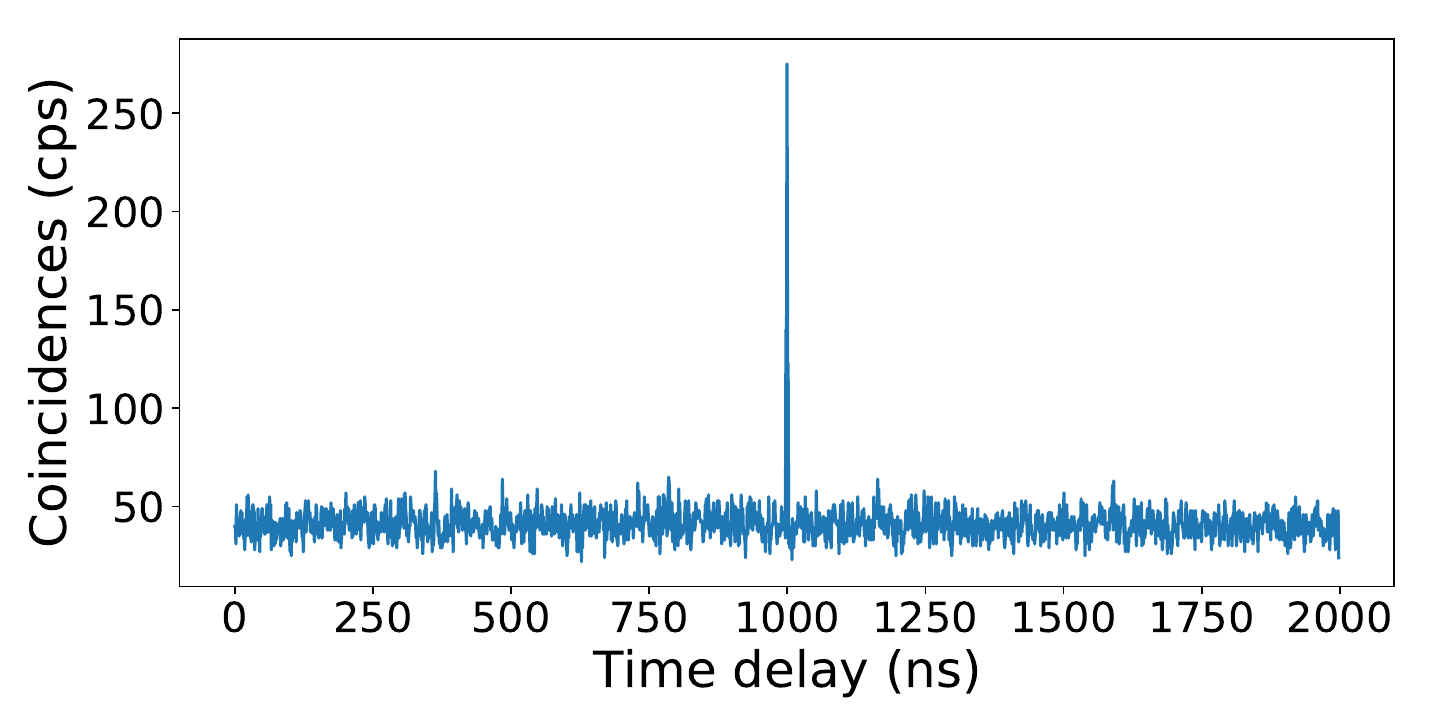}
	\caption{Rough synchronization histogram calculated during the initialization of the protocol to find the time offset with a precision of 2\,ns. To get the time offset, the required delay to set the peak at the $1000^{th}$ bin is measured.}
	\label{fig_firstpeak}
    \end{center}
\end{figure}

At the start of the QKD protocol, the TDCs begin to receive detection events, translated into a list of timestamps for Alice [A(t)] and Bob [B(t)] described in eq.\eqref{timestamp}, with $i$ being the event number among the $N$ registered events :

\begin{equation}
    \left[A(t)\right]=\sum_{i}^{N} \delta(t-t_{a,i}),\quad B(t)=\sum_{j} \delta(t-t_{b,j}).
    \label{timestamp}
\end{equation}

The position of the central coincidence peak (or time offset) $\delta t_0$ can be measured by computing the cross-correlation histogram, as shown in Fig.\ref{fig_3peaks}.
Also, we showed earlier that in order for the protocol to work correctly, the position of the peak must be ultimately known with a precision better than $80$\,ps, to center the coincidence window around it.
However, the complexity of computing a histogram with small bins over a long period of time becomes too high for regular computers.
more specifically, computing a coincidence histogram with bins of $80$\,ps over $1$ second would require more than $10^{11}$ operations, making its computation impractical for commercial computer.
To overcome this issue, we choose to perform a rough synchronization step before computing the cross-correlation function.

When starting the QKD protocol, the pump laser is blocked, and the signal received by the TDCs is only noise, mainly due to dark counts in the detectors, around a few hundreds of events per second.
When unblocking the laser light, the source starts to generate pairs of photons that are subsequently separated and sent simultaneously to Alice and Bob.
On each side, an algorithmic filtering function verifies, every millisecond, if the number of detection is largely greater than the dark-count rate of the detectors.
When the detection rate becomes large enough, the two TDCs prepare a block of timetag acquisitions of $\pm50$\,ms, around the trigger time.
Since the light is unblocked simultaneously for both sides, the coincidence peak can only be in this $100$\,ms time window, which sets a first estimation of the delay $\delta t_{0}$ between correlated events.

Then, the $100$\,ms timetag block is sent to a cross correlation program, plotting the coincidence histogram with bins of $1$\,ns (see Fig.\ref{fig_firstpeak}).
The bin size allows to measure the position of the coincidence peak with a precision of $2$\,ns, which increases again the precision of the measurement of $\delta t_0$.
Finally, a last histogram is plotted over $10$\,ns with bins of $4$\,ps to get our optimal precision for the position of the peak.
At this point, the calculation of the histogram becomes impractical with usual methods.
Usually, a convolution is performed by multiplying the Fourier transform of each function.
The calculation presents a complexity $C_{convolution}=a\log a$, with $a$, the acquisition time divided by the size of the bin used in the histogram.
If we plot the last histogram with a $1$\,s integration time, we have $a=2.5*10^{11}$ leading to $C_{convolution}=\mathcal{O}(10^{12})$ which would take far more than the few seconds we can allow for this calculation.
Another method would be to compare directly each timestamp, leading to a complexity of $C_{timestamp}=N_{A}*N_{B}$, with $N_{A}$ the number of detection from Alice during the acquisition time, and $N_{B}$ the number of detection from Bob in the histogram range during the acquisition time.
The last histogram contains $2500$ bins, and we get on average, $3$\,MHz of detection at Bob's station, and $10$\,MHz at Alice's ($N_{A}=10^{7}$ and $N_{B}=3.10^{6}*4*10^{-12}*2500=3.10^{-2}$.
This means that the complexity of the histogram calculation would be $C_{timstamp}=\mathcal{O}(10^{5})$.
With this, computing a  histogram in less than a second and can be done easily by any computer.
This allows to plot a similar histogram several times per second to compensate for the drift of the clocks, as explained in the next section.

\section{Frequency drift correction}

\begin{figure}[htb]
	\begin{center}
   \includegraphics[width=0.95\columnwidth]{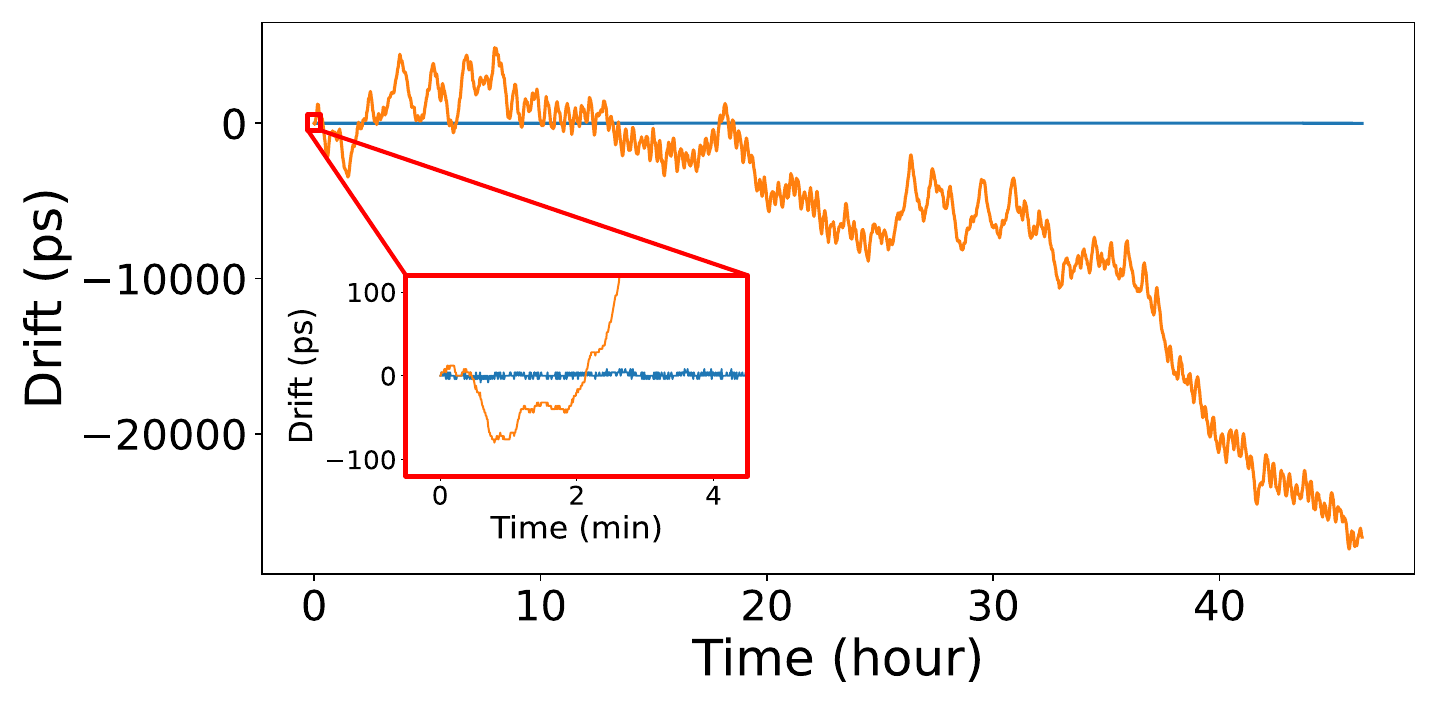}
	\caption{Corrected drift (blue) for a $48$-hour-long measurement, with a variance of $9\,$ps, compared to the drift without correction (orange), drifting up to $27\,$ns during this measurement.}
	\label{fig_drift_cumulated}
    \end{center}
\end{figure}

After measuring the initialization time comes the second challenge of our synchronization protocol, \textit{i.e.} compensating the relative drift of the clocks.
Our setup uses two remotely placed atomic clocks sending pulses at $10$\,MHz to the TDCs, with a precision of $\pm 2.10^{-7}$\,MHz.
This uncertainty on the clock frequencies induces a random temporal drift between the two users, as plotted in Fig.\ref{fig_drift_cumulated} (orange curve).
It shows that the coincidence peak would drift for more than 240\,ps for about 10\,s, setting the coincidence window totally out of the coincidence peak.

To avoid this effect, the drift has to be actively compensated.
It can be achieved either by externally synchronizing the two clocks, or by periodically  measuring the timeshift and post-correcting the drift of the clocks accordingly.
While the first solution requires dedicated hardware and the transmission of classical signals between the users, the second can be ensured by only exploiting the photon pairs generated for the QKD process.
It requires, however, a certain natural stability from the clocks, on the order of few picoseconds per second, which is easily meet using standard rubidium clocks, allowing for re-synchronization every few second.
We choose the second solution and implement it by measuring regularly the central peak position on the coincidence histogram shown in Fig.\ref{fig_3peaks}.

The frequency at which such a measurement must be done depends on the stability of each user's clock. 
For example, the built-in clocks of our TDCs present an average drift of $100\,$ns/s, forcing an estimation of the drift at least every millisecond to guarantee a stabilization under $100$\,ps, which is too demanding for being implemented on standard computers.
Instead, we replace them by rubidium atomic clocks (Spectratime LNRClock 1500), with an average time drift of $7$\,ps/s.
They can therefore drift for $\sim3$\,s without correction, while maintaining the peak mostly inside the coincidence window.
With this, only one measurement of $\delta t_{clock}$ is required every few seconds, which is easily done by our post-treatment program.

The measurement of $\delta t_{clock}$ is done in the same way as the one described at the end of section \ref{sec_offset}, by plotting a histogram over $1$\,ns, with $4$\,ps bins, that can be calculated several times per second.
These calculations allow the extraction of the temporal evolution of $\delta t_{clocks}$, which can then be linked to the frequency drift between the two clocks, to correct it.
This frequency correction allows to minimize the drift between two consecutive correlation measurements, making the stabilization more precise, and resilient to slow external perturbations, such as the variation of the optical path between night and day or the temperature changes at the end stations affecting the atomic clock frequency.

\section{Results}

We test the synchronization protocol on the quantum communication testbed described in section \ref{sec_testbed} for several runs of $48$\,hours each (see ref\cite{QUANTUM} for more QKD-related details).
We achieve a long term synchronization for the whole duration of the QKD protocol, with the clocks never drifting apart from more than $12$\,ps during the whole experiment.
As shown in Fig.\ref{fig_drift_cumulated} (blue line), for a single run, the drift keeps a mean value of $0.08$\,ps and a variance of $9$\,ps.
With these numbers, we keep completely the $80$\,ps wide coincidence peak in the $120$\,ps wide coincidence window at all time.

\begin{figure}[htb]
	\begin{center}
   \includegraphics[width=0.95\columnwidth]{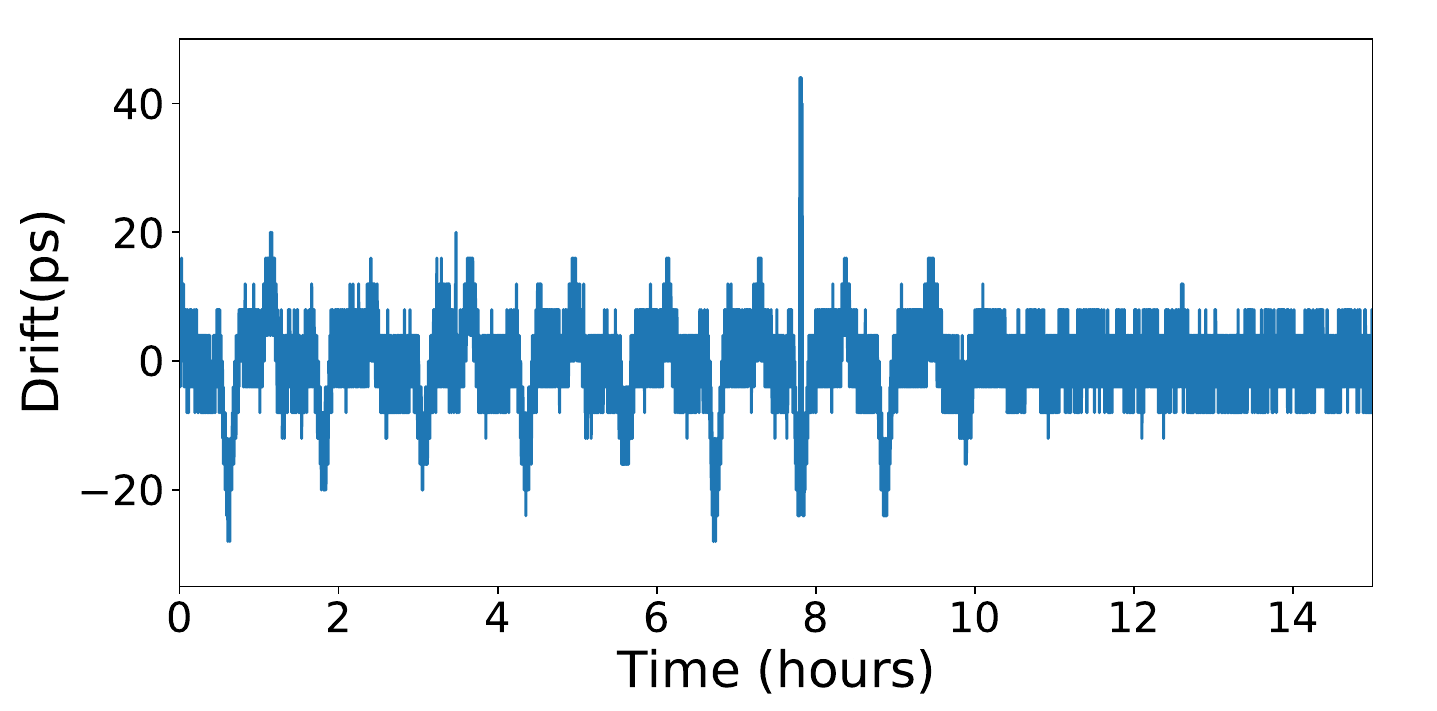}
	\caption{Relative drift between the two remote clocks. An external perturbation destabilized Bob's clock during the first 10 hours of measurement, while the last 5 hours show standard performance for our protocol.}
	\label{fig_drift}
    \end{center}
\end{figure}

We also demonstrate with Fig.\ref{fig_drift} that this implementation is resilient to external perturbations.
During the first 10 hours of the measurement, an external magnetic perturbation affected the frequency of the clock on Bob's side, making its passive stability 20 times worse than usual.
Despite the perturbation, the average drift stays close to zero and the maximum drift between two measurement stays under $44$\,ps, keeping the system suitable for our QKD application.
This perturbation is a typical example of what can happen in an environment for deployed QKD experiments, whose nodes will likely be positioned in less controlled environment than in a laboratory.

Our scheme is also strongly relies on the coincidence rate, i.e. one has to accumulate enough data to perform accurate estimation of the central peak position.
We therefore characterized our synchronization protocol with more losses along the quantum channel.
The synchronization holds with similar characteristics from 0 to up to $32$\,db of total transmission losses (with $19$\,dB being the losses for the deployed system). 
For more than $32$\,dB, the protocol needs more time ($<60$\,s) to accumulate data to create the histogram. With this, the synchronization holds with lower precision (worse than $\pm80$\,ps), which can affect the QKD process.

To overcome this limitation, dedicated photon pairs could be sent in the link, to reduce the time required to integrate a coincidence peak. Practically, we filter the photons generated by the source to send to each user one ITU-DWDM channel for the key distribution process and another one for the synchronization.
In this situation, sending a second channel, dedicated to synchronization, would effectively double the amount of data sent, therefore allowing for the addition of $3$\,dB of losses on the link.
Our source can generate up to $40$ pairs of $100$`\,GHz channels, which could allow potentially to push the $32$\,dB limitation to $48$\,dB, while maintaining the time precision achieved in this paper.

\section{Conclusion}

In conclusion, we have successfully implemented a time synchronization protocol on a real-field QKD link, pacing two rubidium atomic clocks together with less than $12$\,ps of error at all time.
Beside being on par with the most recent state-of-the-art protocols \cite{lipinski2018white, valivarthi2022picosecond, merlo2022wireless}, our implementation uses only components already implemented in entanglement-based QKD systems, and therefore is particularly adapted for QKD time synchronization.
Furthermore, since the photon pairs are directly used for synchronization, this protocol also passively corrects the optical path variations induced by the deployed fibers.
In our implementation, it allows to maintain at all time our coincidence peak inside of our coincidence window, and therefore continuously guarantees an optimal secret key rate.
Lastly, let us mention that this synchronization protocol is not limited to applications in quantum technologies, and could be deployed on any system, simply using an entangled photon pair source, two single photon detectors and two TDCs.
Data are available from the authors on reasonable request.

\subsection*{Acknowledgements}
This work has been conducted within the framework of the French government financial support managed by the Agence Nationale de la Recherche (ANR), within its Investments for the Future programme, under the Université Côte d’Azur UCA-JEDI project (Quantum@UCA, ANR-15-IDEX-01), and under the Stratégie Nationale Quantique through the PEPR QCOMMTESTBED project (ANR 22-PETQ-0011). This work has also been conducted within the framework of the OPTIMAL project, funded by the European Union and the Conseil Régional SUD-PACA by means of the ‘Fonds Européens de développement regional’ (FEDER). The authors also acknowledge financial support from the European comission, project 101114043-QSNP, and project 101091675-FranceQCI. The authors are grateful to S. Canard, A. Ouorou, L. Chotard, L. Londeix and Orange for the installation and the connection of the dark fibers between the three different sites of our network, as well as for all the support they provided for their characterization. The authors also thank the Métropole Nice Côte d’Azur and the Inria Centre at Université Côte d’Azur for the access to their buildings and for their continuous help in making this network a reality. The authors also acknowledge IDQuantique and Swabian Instruments GmbH teams for all the technical support and the development of new features that were needed for the implementation of our operational QKD system and related experiment.

\nocite{*}
\bibliographystyle{apsrev4-1}
\bibliography{Synchronization}

\end{document}